\documentclass[a4paper,12pt]{article}
\usepackage{amsmath}
\usepackage{amssymb}
\usepackage{latexsym}
\usepackage{graphicx}
\newcommand{\be}{\begin{eqnarray}}
\newcommand{\ee}{\end{eqnarray}}
\begin{document}
\begin{titlepage}

\begin{centering}
\vspace{.3in}
{\Large{\bf M\o ller's Energy-Momentum Complex for a Spacetime Geometry on a Noncommutative Curved D3-Brane}}
\\

\vspace{.5in} {\bf I. Radinschi$^*$\footnote{radinschi@yahoo.com}, Th. Grammenos$^\dag$\footnote{thgramme@uth.gr}},\\

\vspace{0.3in}
{\it $^*$ Department of Physics, ``Gh. Asachi'' Technical University,\\
Iasi, 700050, Romania\\
$^\dag$ Department of Mechanical \& Industrial Engineering, University of Thessaly,\\
383 34 Volos, Greece\\
}
\end{centering}

\begin{abstract}
\noindent M\o ller's energy-momentum complex is employed in order to determine the energy and
momentum distributions for a spacetime described by a ``generalized Schwarzschild" geometry in
(3+1)-dimensions on a noncommutative curved D3-brane in an effective, open bosonic string theory.
The geometry considered is obtained by an effective theory of gravity coupled with a nonlinear
electromagnetic field and depends only on the generalized (effective) mass and charge which incorporate
corrections of first order in the noncommutativity parameter.
\end{abstract}
\end{titlepage}

\section{Introduction}

The localization of energy and momentum in curved spacetime has been one of the most intricate issues and a subject
of extensive research since the outset of general relativity.  There has been a large number of attempts aiming at
the localization of  energy-momentum in this context, and they all have resulted in various (pseudotensorial)
energy-momentum complexes. In order to solve the problem of energy and momentum localization, Einstein \cite{Einstein} was the first
who proposed such a prescription followed, mainly, by Tolman \cite{Tolman}, Papapetrou \cite{Papapetrou}, Landau and Lifshitz \cite{Landau},
Bergmann and Thomson \cite{Bergmann}, Goldberg \cite{Goldberg}, M\o ller \cite{Moller} and Weinberg \cite{Weinberg}.
With the exception of M\o ller's prescription,
who proposed a new expression for an energy-momentum complex that can be utilized in any coordinate system,
all the aforementioned energy-momentum complexes are restricted to quasi-Cartesian coordinates. However,
one should point out that, from the very beginning, the use of energy-momentum complexes has been heavily
criticized on the grounds of their nontensorial character which could lead to rather dubious physical interpretations \cite{Chandra}.
Furthermore, different localization prescriptions could lead to different energy distributions for the same spacetime
geometry \cite{Bergqvist}, while due to their local character it was widely accepted that the total energy-momentum of the
gravitational field could not be localized \cite{Chen}. In 1973, Misner, Thorne and Wheeler \cite{Misner} sustained that ``anybody
who looks for a magic formula for local gravitational energy-momentum is looking for the right answer to
the wrong question". This statement gives the actual meaning of the nonuniqueness of the pseudotensor for
the energy-momentum. However, they concluded that the energy is indeed localizable only for spherical systems.
In 1978, Cooperstock and Sarracino \cite{Sarracino} showed that if the energy is localizable in spherical systems, then
it is also localizable in any spacetime. In 1990, Bondi \cite{Bondi} maintained that ``a non-localizable form of energy
is not admissible in general relativity, because any form of energy contributes to gravitation and so its
location can in principle be found". Further, worth mentioning attempts to deal with the problems stemming
from the concept of energy-momentum complex rely on the quasi-local energy approach \cite{Brown}, while there has
been also a considerable amount of investigations on the concept of superenergy and the Bel-Robinson tensor \cite{Senovilla}.
The issue of the application of the pseudotensorial energy-momentum complexes for the localization of energy
was almost abandoned until 1990, when Virbhadra and other scientists revived and enlightened the subject \cite{Virbhadra}.
Since then a growing number of studies where different energy-momentum complexes are applied to a plethora of
spacetime geometries has been published \cite{Garecki}. In 1996, Aguirregabiria, Chamorro and Virbhadra \cite{Aguirregabiria} established
an important result showing that the Einstein, Tolman Landau-Lifshitz, Papapetrou and Weinberg (henceforth ETLLPW)
energy-momentum complexes essentially coincide if calculations are carried out in Kerr-Schild Cartesian coordinates
for a general metric of the Kerr-Schild class. Furthermore, the results of the ETLLPW energy-momentum complexes
were found to be identical with the older results of Penrose \cite{Penrose} and Tod \cite{Tod}, who had applied the notion of
quasi-local mass. In 1999, in an attempt to support and stress the importance and usefulness of the concept of
energy-momentum complex for the localization of energy, Chang, Nester and Chen \cite{Chang} showed that there exists a
direct relationship of every energy-momentum complex with a Hamiltonian boundary term. Consequently, an
energy-momentum complex provides a quasi-local, legitimate and acceptable expression. Work on this topic is still going on.

At this point it is worth noting several interesting results obtained in the last years in the study
of the problem of energy localization in $2 + 1$ and $1 + 1$ dimensions by using energy-momentum complexes \cite{Vagenas}.
Further, a great deal of work has been done in the context of the tele-parallel version of gravity,
where it is demonstrated that the energy-momentum complexes yield the same results as their tele-parallel versions \cite{Nashed}.

In recent years, many physically satisfactory results were obtained by applying M\o ller's prescription \cite{Aygun}.
In this paper, we wish to emphasize the importance of the localization of energy using the same prescription,
i.e., M\o ller's energy-momentum complex. To this purpose, we extend our earlier study \cite{Grammenos} and apply M\o ller's
prescription to evaluate the energy and momentum distributions for a metric describing a generalized
Schwarzschild geometry in $(3+1)$ dimensions on a noncommutative D3-brane in the context of an effective,
open bosonic string theory. The rest of the paper is organized as follows: In section 2 the metric of
the generalized Schwarzschild geometry and the general lines of its derivation are presented. Section 3
contains the definition of M\o ller's energy-momentum complex, while in section 4 we give the details of the
calculations for the energy and momentum distributions. Finally, section 5 is devoted to a summary of the
results obtained and to conclusions. The calculations have been performed using {\it Mathematica} and {\it Maple},
the latter having attached the GrTensor platform.

Conventions: Throughout this paper we have used geometrized units $(c = 1, G = 1)$, the metric's
Lorentzian signature is $(-, +, +, +)$, and Greek (Latin) indices are running from 0 to 3 (1 to 3).

\section{The Generalized Schwarzschild Geometry}
The nonperturbative approach to string theory has led to certain higher dimensional,
extended objects called Dp-branes and defined as spacetime hypersurfaces onto which open
strings can be attached and Dirichlet boundary conditions can be formulated \cite{Johnson}. At low
energy scales or for slowly varying fields, the dynamics of Dp-branes is controlled by the
effective Dirac-Born-Infeld (DBI) action, a natural generalization of the Dirac-Nambu-Goto action.
The DBI action, obtained by one of the simplest nonpolynomial Lagrangeans invariant under
reparametrizations of the world-volume, is based on a U(1) gauge field ``living'' on the world-volume
swept out by the Dp-brane coupled to the inherent geometry of the world-volume.

In order to address the Hawking radiation phenomenon, the
information loss paradox and, if possible, to formulate an
effective theory of quantum gravity at the Planck scale, Kar and
Majumdar \cite{Kar1} utilized the DBI-framework of a brane
world-volume incorporating Einstein's theory of general relativity
coupled to Born-Infeld nonlinear electrodynamics. Then,
considering the evolution of gravity on a noncommutative, curved
D3-brane in an effective, open bosonic string theory setting (by
using the Seiberg-Witten map \cite{Seiberg}, the authors
transformed the ordinary gauge dynamics to a noncommutative U(1)
gauge dynamics on the D3-brane) they obtained a generalized
Schwarzschild geometry at the Planck scale. Essentially, in going
towards the Planck regime, the authors combined an effective
theory of gravity with a U(1) noncommutative gauge field and
managed to proceed from a Schwarzschild-like geometry (described
by eq.(11) in our previous paper \cite{Grammenos}), obtained with
an ordinary spacetime in the classical regime, to a generalized
Schwarzschild geometry on a noncommutative D3-brane, whereby a
static gauge condition was used in order to incorporate a
nontrivial induced metric on the brane world-volume. From a
topological point of view, the generalized Schwarzschild
geometry obtained describes a Euclidean manifold $\mathbb{R}^2\times S^2$.

Thus, following Kar and Majumdar \cite{Kar1}, we consider a Euclidean world-volume spanned
by $(x_1,x_2,x_3,x_4)$ with a signature $(+,+,+,+)$, where the Lorentzian signature may be
obtained by an analytic continuation $x_4\rightarrow it$, and the noncommutative D3-brane dynamics
is obtained from the, appropriately modified, effective DBI-action given by
\begin{equation}\label{1}
\hat{S}_\text{DBI}=T_{D}^{nc}\int d^4 x (\sqrt{G} -\sqrt{G+2\pi \alpha^{\prime}\hat{F}}),
\end{equation}
where $T_{D}^{nc}$ is the noncommutative D3-brane tension,  $G\equiv \mbox{det} G_{\mu\nu}$, where
$G_{\mu\nu} = g_{\mu\nu} -(Bg^{-1}B)_{\mu\nu}$ is the effective metric on the brane, $B_{\mu\nu}$ is a constant 2-form induced on
the world-volume of the brane, and the noncommutative U(1) field strength on the world-volume is given by
\begin{equation}\label{2}
\hat{F}_{\mu\nu}=\partial_{\mu}\hat{A}_{\nu}-\partial_{\nu}\hat{A}_{\mu}
+\Theta^{\rho\lambda}\partial_{\rho}\hat{A}_{\mu}(x)\partial_{\lambda}\hat{A}_{\nu}(y)+O(\Theta^2),
\end{equation}
where $\hat{A}_{\mu}$ is the noncommutative gauge potential and $\Theta$ is the noncommutativity parameter
expressing the noncommutative corrections in the theory.
Now, the action for a curved D3-brane becomes
\begin{equation}\label{3}
\hat{S}=\frac{1}{16\pi}\int d^4 x \sqrt{G}R+\hat{S}_\text{DBI}
\end{equation}
with $R$ the scalar curvature, while the appearing (string) gauge
field corrections described by the higher order terms can vanish
identically in a special combination of the gauge field. Hence,
the effective noncommutative string action is finally obtained as
\begin{equation}\label{4}
\hat{S}=\int d^4 x
\sqrt{G}\left(\frac{1}{16\pi}R-\frac{1}{4}G^{\mu\lambda}G^{\nu\rho}\hat{F}_{\mu\nu}\ast\hat{F}_{\lambda\rho}\right),
\end{equation}
where the star $\ast$ denotes the Moyal product.

With a specific gauge choice for the effective metric
$G_{\mu\nu}$, the Schwarzschild-like geometry (eq.(11) in \cite{Grammenos})
can be generalized to the effective noncommutative theory (\ref{4}) at
the Planck scale. In other words, a generalized semi-classical
solution on a noncommutative curved D3-brane with the action (\ref{3})
can be constructed using the effective metric $G_{\mu\nu}$. In the presence
of a nonzero Neveu-Schwarz B-field (being equivalent to a constant
magnetic field on the brane), the effective metric for the
aforementioned geometry is finally given in spherical-polar
coordinates by the following generalized Schwarzschild line
element \cite{Kar1}:
\begin{equation}\label{5}
\begin{split}
ds^2 = &-\left(1-\frac{2M_\textit{eff}}{r}-\frac{r_{c}^{4}}{r^4}+\frac{2M_\textit{eff}\, r_{c}^{4}}{r^5}\right)dt^2
+\left(1-\frac{2M_\textit{eff}}{r}-\frac{r_{c}^{4}}{r^4}+\frac{2M_\textit{eff}\, r_{c}^{4}}{r^5}\right)^{-1}dr^2\\
&+ \left(1-\frac{r_{c}^{4}}{r^4}\right)r^2 d\theta^2+\left(1-\frac{r_{c}^{4}}{r^4}\right)^{-1}r^2\sin^2\theta \, d\phi^2
\end{split}
\end{equation}
with the effective parameters of the noncommutative formalism \cite{Kar2}
\begin{equation}\label{6}
M_\textit{eff}=M\left(1-\frac{\Theta}{r^2}+O(\Theta^2)+\ldots\right)
\end{equation}
\begin{equation}\label{7}
r_{c}^{2}=Q\left(1-\frac{\Theta}{2r^2}+O(\Theta^2)+\ldots\right),
\end{equation}
where $M_\textit{eff}$  is the generalized ADM mass,  $r_{c}^{2}$ acts as an effective charge $Q_\textit{eff}$, and $Q$ is the
electric charge. Keeping only first order terms in the noncommutativity parameter $\Theta$,
the line element (\ref{5}) becomes:
\begin{equation}\label{lineelement}
\begin{split}
ds^2 = &-\left[1-\frac{2M}{r}\left(1-\frac{\Theta}{r^2}\right)-\frac{Q^2}{r^4}\left(1-\frac{\Theta}{r^2}\right)
+\frac{2M\,Q^2}{r^5}\left(1-\frac{2\Theta}{r^2}\right)\right]dt^2
\\
& + \left[1-\frac{2M}{r}\left(1-\frac{\Theta}{r^2}\right)-\frac{Q^2}{r^4}\left(1-\frac{\Theta}{r^2}\right)
+\frac{2M\,Q^2}{r^5}\left(1-\frac{2\Theta}{r^2}\right)\right]^{-1}dr^2\\
& +\left[1-\frac{Q^2}{r^4}\left(1-\frac{\Theta}{r^2}\right)\right]r^2 d\theta^2
+\left[1-\frac{Q^2}{r^4}\left(1-\frac{\Theta}{r^2}\right)\right]^{-1}r^2\sin^2\theta \, d\phi^2.
\end{split}
\end{equation}
As it is evident, the geometry is characterized only by the effective mass
and the effective charge parameters. It is to be noted that the effective four-dimensional
spacetime described by the line element (\ref{5}) may be obtained from a five-dimensional gravity
with a small scale along the fifth dimension \cite{Kar2}. The effective theory solution described by
the line element (\ref{5}) breaks the spherical symmetry, it is of Petrov type I and describes a
$G_2$ space (two Killing vectors) which is not maximally symmetric. Further, it has an obvious
curvature singularity at $r = 0$ and two coordinate singularities  at $r = r_c$ and $r = 2M_\textit{eff}$.
In fact, the noncommutative corrections may be viewed as a stretch in the event horizon of
the object having this unusual exterior geometry.

In the classical regime, the corrections in the effective metric are small and may be ignored.
Thus, in this case, the generalized geometry described by (\ref{lineelement}) is reduced to the classical
Schwarzschild spacetime geometry.

\section{M\o ller's Energy-Momentum Complex}
Arguing that the singling out of quasi-Cartesian coordinates in the Einstein energy-momentum
complex is not satisfactory from the viewpoint of general relativity, M\o ller \cite{Moller} searched for an
expression for energy and momentum that would be independent from any particular coordinate system.
To that purpose, M\o ller seeked for a quantity with identically vanishing divergence that could be
added to Einstein's energy-momentum complex so that their sum could be transformed as a tensor for
spatial transformations. M\o ller's quantity should be form-invariant and dependent on the metric
tensor as well as on its first and second derivative. Further, in linear transformations it
should behave as a tensor density satisfying certain conditions (see, e.g., \cite{Xulu} for details).
Employing this procedure, M\o ller arrived at the following energy-momentum complex in a four-dimensional background:
\begin{equation}\label{8}
J^{\mu}_{\nu}=\frac{1}{8\pi}\xi_{\nu,\lambda}^{\mu\lambda}
\end{equation}
where M\o ller's superpotential $\xi_{\nu}^{\mu\lambda}$ has the form
\begin{equation}\label{9}
\xi_{\nu}^{\mu\lambda}=\sqrt{-g}
\left(\frac{\partial g_{\nu\sigma}}{\partial x^{\kappa}}-\frac{\partial g_{\nu\kappa}}{\partial x^{\sigma}}\right)g^{\mu\kappa}g^{\lambda\sigma}
\end{equation}
with the antisymmetry property
\begin{equation}\label{10}
\xi_{\nu}^{\mu\lambda}=-\xi_{\nu}^{\lambda\mu}.
\end{equation}
It is easily seen that M\o ller's energy-momentum complex satisfies the local conservation equation
\begin{equation}\label{11}
\frac{\partial J_{\nu}^{\mu}}{\partial x^{\mu}}=0,
\end{equation}
where $J_{0}^{0}$ gives the energy density and $J_{i}^{0}$ gives the momentum density components.

Finally, in M\o ller's prescription the energy and momentum in a four-dimensional background are obtained by
\begin{equation}\label{12}
P_{\nu}=\iiint J_{\nu}^{0} dx^{1}\, dx^{2} \, dx^{3}.
\end{equation}
Specifically, the energy of a physical system is given by
\begin{equation}\label{13}
E=\iiint J_{0}^{0} dx^{1}\, dx^{2} \, dx^{3}.
\end{equation}
As it was pointed out in the beginning of this section, the calculations are not anymore restricted
to quasi-Cartesian coordinates, but can be utilized in any coordinate system. Indeed, Lessner \cite{Lessner}
has concluded that M\o ller's energy-momentum complex is a powerful expression for the localization
of energy and momentum in general relativity.

\section{Energy and Momentum Density Distributions}
For the metric under consideration given by the line element (\ref{lineelement}), we obtain eight non-vanishing M\o ller superpotentials,
with the antisymmetric property (\ref{10}) taken into account:
\begin{eqnarray}
\xi_{2}^{21}=-\xi_{2}^{12}&=&-\frac{2\sin\theta[r^6+Q^2(r^2-2\Theta)]}{r^6[r^6+Q^2(-r^2+\Theta)]}
\left\{r^7+Q^2(-r^{3}+r\Theta)\right.\nonumber\\
&&\left.+2M[-r^6+Q^2(r^2-2\Theta)+r^4\Theta]\right\}\label{14}\\
\nonumber \\
\xi_{3}^{31}=-\xi_{3}^{13} &=& -\frac{2\sin\theta[r^6+Q^2(-3r^2+4\Theta)]}{r^6[r^6+Q^2(-r^2+\Theta)]}
\left\{r^7+Q^2(-r^{3}+r\Theta)\right.\nonumber\\
&&\left.+2M[-r^6+Q^2(r^2-2\Theta)+r^4\Theta]\right\}\label{15}\\
\nonumber \\
\xi_{3}^{32}=-\xi_{3}^{23}&=& \frac{2r^6}{-r^6 +Q^2 r^2 -\Theta Q^2}\cos\theta\label{16}\\
\nonumber\\
\xi_{0}^{01}=-\xi_{0}^{10}&=&\frac{-2\sin\theta}{r^6}\left[r^2(2Q^2r+Mr^4-5MQ^2)\right.\nonumber\\
& &\left.+ \Theta(-3Q^2r-3Mr^4+14MQ^2)\right]\label{17}
\end{eqnarray}
By substituting the superpotentials given by eqs.(\ref{14}-\ref{17}) into eq.(\ref{8}) we obtain the energy density distribution
\begin{equation}\label{18}
\begin{split}
J_0^0 =\frac{\sin\theta}{4\pi r^7} &\left[\Theta(-15Q^2r-6Mr^4+84MQ^2)\right.\\
&\left.+2Q^2r^2(3r-10M)\right],
\end{split}
\end{equation}
while all the momentum density distributions vanish:
\begin{equation}\label{19}
J_{i}^{0}=0.
\end{equation}
Thus, all the momentum components  are zero.
Inserting eq.(\ref{18}) into eq.(\ref{13}) and evaluating the integral, we obtain the energy contained in a ``sphere'' of radius R:
\begin{equation}\label{20}
\begin{split}
E=M-\frac{1}{R^{6}} &\{R^2 Q^2(2R-5M) \\
&- \Theta [3R^4 M+Q^2(3R-14M)]\}.
\end{split}
\end{equation}
Evidently, the energy contained in this ``sphere'' depends only on the radius $R$, the mass $M$ and the charge $Q$
generalized by the noncommutativity parameter $\Theta$ and given by equations (\ref{6}-\ref{7}).
In the zero-charge limit $Q\rightarrow 0$, one obtains from (\ref{20}):
\begin{equation}\label{21}
E=M\left(1+\frac{3\Theta}{R^2}\right),
\end{equation}
while, in the case there are not any noncommutativity corrections, i.e. $\Theta = 0$, (\ref{20}) yields
\begin{equation}\label{22}
E=M-\frac{2Q^2}{R^3}\left(1-\frac{5}{2}\frac{M}{R}\right).
\end{equation}
Taking the asymptotic limit $R\rightarrow \infty$, one derives from (\ref{20}) that the energy equals the (ADM) mass $M$:
\begin{equation}\label{23}
E=M,
\end{equation}
a result that can also be obtained from (\ref{20}) when $Q$ and $\Theta$ vanish simultaneously.
However, one must point out that the energy $E$ equals the (ADM) mass $M$ also when $\Theta = 0$ and $R=\frac{5}{2}M$.
No reasonable explanation could be found for this result that was obtained also in \cite{Grammenos}.
An explanation of this rather unexpected result, namely that the energy $E$ equals the (ADM) mass $M$ for
a finite radius $R$, could be possibly offered by a detailed study of the properties
of the spacetime geometry described by the line element (\ref{5}).

\section{Conclusions}
The different attempts made to develop a generally accepted expression for the energy-momentum
density yield a plethora of energy-momentum definitions. Indeed, for solving the problem of
energy and momentum localization many attempts have been made in the past, but it still
remains an important open issue to be settled. Even though the energy-momentum complexes
are apparently useful for the localization of energy, there are still doubts that these prescriptions
could give acceptable results for any given space-time for reasons discussed in section 1.
This is, among other things, due to the fact that many of the energy-momentum complexes
(e.g., ELLTPW and Bergmann-Thomson) are restricted to the use of particular coordinate
systems and only the M\o ller energy-momentum complex allows one to calculate the energy
distribution in any coordinate system. Nevertheless, the results obtained by several
authors (\cite{Garecki}-\cite{Grammenos}) have demonstrated that the energy-momentum complexes
are useful concepts for evaluating the energy and momentum distributions in general relativity.

In this work, the energy and momentum distributions for a semiclassical generalized Schwarzschild
solution in $(3+1)$ dimensions that reduces to the classical Schwarzschild solution in the classical
regime and is derived by an effective, open bosonic string theory on a noncommutative curved D3-brane
in a static gauge, with a mass and charge being noncommutatively corrected to first order in the
noncommutativity parameter $\Theta$ (giving a generalized mass and a generalized charge),
have been calculated by using M\o ller's energy-momentum complex. There are totally eight nonvanishing
M\o ller superpotentials and all momentum components are found to be zero. The total energy contained
in a ``sphere'' is computed, while its value is obtained in the zero-charge limit, in the asymptotic
limit, and when there are no noncommutativity corrections.

Finally, an investigation of the considered four-dimensional exterior geometry described by
the line element (\ref{5}) showed that it is of Petrov type I. However, one must point out
that this classification might change by some proper combination of mass and charge.
Further, the metric describes a G2 space (with $\displaystyle{\frac{\partial}{\partial t}}$
and $\displaystyle{\frac{\partial}{\partial \phi}}$ as Killing vectors) which is not maximally
symmetric (non-zero Weyl projective tensor) and it is not an Einstein space. A calculation
and examination of the curvature invariants for this metric, even though computationally
tedious, would be of interest for a clarification of further properties this spacetime geometry might have.

\section*{Acknowledgements}
The authors are indebted to Dr. E.C. Vagenas for useful suggestions and valuable comments.

\end{document}